\documentclass[aps,prd,twocolumn,showpacs,amsmath,amssymb]{revtex4-1}
\usepackage{graphicx}
\usepackage{subfig}
\usepackage{caption}
\usepackage{epstopdf}
\usepackage{color}
\usepackage{multirow}
\usepackage{setspace}
\usepackage{overpic}
\usepackage{amsmath}
\usepackage{amssymb}
\usepackage{booktabs}
\usepackage{lineno}
\usepackage{bm}
\usepackage{rotating}
\usepackage{verbatim}
\usepackage[utf8]{inputenc}
\graphicspath{{Fig/}}
\usepackage{bigstrut}
\usepackage{hyperref}
\usepackage{natbib}
\begin{document}


\title{\boldmath Mini-review of charmonium weak decays at BESIII}

\author{Xu-Ze Li}
\affiliation{School of Physics, Sun Yat-sen University, Guangzhou 510275, China}
\author{Kai-Xin Fan}
\affiliation{School of Physics, Sun Yat-sen University, Guangzhou 510275, China}
\author{Zheng-Yun You}
\email[Zheng-Yun You, ]{youzhy5@mail.sysu.edu.cn}
\affiliation{School of Physics, Sun Yat-sen University, Guangzhou 510275, China}
\author{Yu Zhang}
\email[Yu Zhang, ]{yuzhang@usc.edu.cn}
\affiliation{University of South China, Hengyang 421001, China}
\author{Minggang Zhao}
\email[Minggang Zhao, ]{zhaomg@nankai.edu.cn}
\affiliation{Nankai University, Tianjin 300071, China}

\begin{abstract}

The weak decays of charmonium states such as $J/\psi$ and $\psi(2S)$ are instrumental in probing both nonperturbative QCD dynamics and the flavor structure of the Standard Model~(SM). The extreme rarity of charmonium weak decays renders them highly sensitive to physics beyond the SM, particularly in channels that are heavily suppressed in the SM, such as flavor-changing neutral-current~(FCNC) decays. This review highlights the critical role of the BESIII experiment, which leverages an unprecedented data sample of over $10^{10}$ $J/\psi$ and $2.7\times10^{9}$ $\psi(2S)$ events to achieve leading sensitivity in searches for charmonium weak decays. We present the latest and most stringent upper limits established by BESIII on various semileptonic, nonleptonic, and FCNC charmonium weak decay channels.

\end{abstract}

\maketitle

\section{Introduction}

Charmonium, the bound state of a charm quark and its anti-quark~($c\bar{c}$), including the states such as $J/\psi$ and $\psi(2S)$, is an ideal system for studying both quantum chromodynamics~(QCD) and the electroweak interaction~\cite{1_rev_1, 2_rev_2,3_theo_1_LQCD_semionly}. Since the masses of $\mathrm{J} / \psi~(3.097~\mathrm{GeV}/c^2)$ and $\psi(2S)~(3.686~\mathrm{GeV}/c^2)$ lie far below the open-charm threshold ($3.73~\mathrm{GeV}/c^2$), they predominantly decay via the Okubo-Zweig-Iizuka suppressed strong or electromagnetic processes, which proceed via the annihilation of the $c\bar{c}$ pair into three gluons or virtual photons. However, the weak decays of charmonium, where one of the constituent quarks decays via a $W$ boson emission or exchange, remain theoretically allowed. Moreover, the narrow total widths resulting from this suppression, combined with the clean experimental environment of $e^+e^-$ colliders, render the search for such rare weak decays feasible.

Charmonium weak decays, despite their extremely small branching fractions~(BFs), serve as a unique, clean laboratory to precisely test the SM and search for new physics~(NP) beyond the standard model~(BSM) for several critical reasons. 

First, they provide crucial SM tests, as they are sensitive to the non-perturbative dynamics of the $c\bar{c}$ bound state. This requires the precise calculation of transition form factors and wave functions, which can be constrained via various methods~\cite{10_theo_8_Spin_symmetry, 4_theo_2_QCDSR, 5_theo_3_CLFQM_2008, 6_theo_4_CLFQM_2024, 8_theo_5_BSW, 7_theo_6_CCQM_semionly, 9_theo_7_BSmethod, 3_theo_1_LQCD_semionly, 11_theo_9_Spin_symmetry, 12_theo_9_Factorization_model, 13_theo_11_QCDSR_non-lep, 14_theo_12_factorization_approximation_non-lep, 15_theo_13_FCNC_QCDSR}. 

Second, the highly suppressed nature of these decays in the SM makes them exquisitely sensitive probes for NP searches. The BSM scenarios, such as super-symmetry~(SUSY), left-right symmetric model, or models addressing the fermion mass hierarchy, could enhance these BFs, particularly in channels involving FCNC decays~\cite{1_rev_1,2_rev_2,16_rev_3, 17_BSM_1}. 

Finally, these searches are driven by a significant experimental opportunity --- the BESIII experiment. The BESIII detector~\cite{Ablikim:2009aa} records symmetric $e^+e^-$ collisions provided by the BEPCII storage ring~\cite{Yu:IPAC2016-TUYA01}
in the center-of-mass energy range from 1.84 to 4.95~GeV, with a peak luminosity of $1.1 \times 10^{33}\;\text{cm}^{-2}\text{s}^{-1}$
achieved at $\sqrt{s} = 3.773\;\text{GeV}$. BESIII has accumulated the world's largest samples of $J/\psi$ and $\psi(2S)$ events produced at rest in $e^+e^-$ annihilation, with $(10087\pm44)\times10^6$ $J/\psi$ events and $(2712.4 \pm 14.3) \times 10^6$ $\psi(2S)$ events, providing unprecedented sensitivity to reach deep into the predicted SM territory and constrain BSM theories~\cite{BESIII-jpsi-Events-1,BESIII-jpsi-Events-2,BESIII-psi2S-Events}. The massive dataset, particularly the $10$ billion $J/\psi$ events , provides the necessary statistical power to search for the extremely rare processes, allowing BESIII to improve upon previous best limits on charmonium weak decay searches~\cite{18_rev_4}.

This review summarizes the theoretical predictions and the recent experimental results from the BESIII collaboration concerning the charmonium weak decays, focusing on $J/\psi$ semileptonic, nonleptonic, and FCNC channels, as well as the results from the $\psi(2S)$ state.

\section{Mechanism of Charmonium Weak Decays}

The weak decay of a charmonium state, such as $J/\psi$ or $\psi(2S)$, proceeds primarily through the decay of one of its constituent quarks, $c \to s/d + W^+$; the $W^+$ subsequently decays into a lepton pair or a quark pair, and the spectator antiquark $\bar{c}$ combines with the resulting $s$ or $d$ quark to form a hadron in the final state~\cite{9_theo_7_BSmethod}.

\subsection{Charmonium semileptonic decays}
\begin{figure}[htbp]
    \centering
    \includegraphics[width=1.0\linewidth]{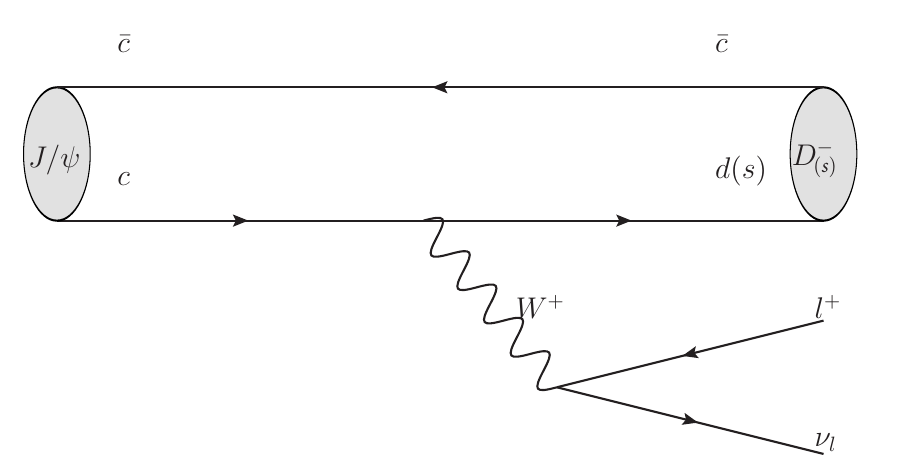}
    \caption{Tree-level Feynman diagram for the charmonium semileptonic decay $J/\psi \to D_{(s)}^{-} l^+ \nu_l$. The diagram was generated using JaxoDraw~\cite{Jaxodraw_ref}.}
    \label{fig:Feynman_Diag_semilep}
\end{figure}

The total width $\Gamma_{Weak}(\psi)$~(where $\psi$ denotes $J/\psi$ or $\psi(2S)$) consists of several components determined by the final state. These include semileptonic decays, such as $\psi \to D_{(s)}^{(*)-} l^+ \nu_l+c.c.$~($l$ denotes $e$ and $\mu$), as shown in Fig.~\ref{fig:Feynman_Diag_semilep}. Throughout this paper, charge-conjugate processes are always implied. 

The charmonium semileptonic decays are governed by tree-level processes mediated by a virtual $W$ boson, resulting from the $c \to (s/d) l^+ \nu_l$ transition. The semileptonic decays of $J/\psi$ and $\psi(2S)$ contain both Cabibbo-suppressed mode~(with a $D^{\pm}$ meson in the final state) and Cabibbo-favored mode~(with a $D_s^\pm$ in the final state). 

Numerous theoretical calculations have been conducted with various QCD frameworks. 
In 1994, M.~A.~Sanchis-Lozano analyzed the weak decays of heavy quarkonium with the heavy quark spin symmetry model~(HQSS)~\cite{10_theo_8_Spin_symmetry}.
In 2007, Y.~M.~Wang et al. studied the transition form factors for semileptonic weak decays of $J/\psi$ in the framework of QCD sum rules~(QCDSR)~\cite{4_theo_2_QCDSR}.
Y.~L.~Shen et al. and Z.~J.~Sun et al. investigated the semileptonic and nonleptonic weak decays of charmonium within the covariant light-front quark model~(CLFQM)~\cite{5_theo_3_CLFQM_2008, 6_theo_4_CLFQM_2024} in 2008 and 2024, respectively.
In 2013, R.~Dhir employed the Bauer, Stech and Wirbel~(BSW) model to estimate the weak decays of heavy quarkonium~\cite{8_theo_5_BSW}.
In 2015, M.~A.~Ivanov et al. investigated the exclusive semileptonic decays $J/\psi \to D_{(s)}^{(*)-} l^+ \nu_l$ in a covariant constituent quark model~(CCQM) with infrared confinement~\cite{7_theo_6_CCQM_semionly}.
In 2016, T.~H.~Wang et al. studied the weak decays of $J/\psi$ using the Bethe–Salpeter~(BS) method~\cite{9_theo_7_BSmethod}. 
In 2024, Y.~Meng et al. performed the first Lattice QCD~(LQCD) calculation on the semileptonic decay of $J/\psi$~\cite{3_theo_1_LQCD_semionly}.

Theoretical predictions for the BFs range from $10^{-12}$ to $10^{-10}$ for $\mathcal{B}(J/\psi \rightarrow D^{-} l^{+} \nu_l)$ and from $10^{-10}$ to $10^{-9}$ for $\mathcal{B}(J/\psi \rightarrow D_{s}^{-} l^{+} \nu_l)$. In most model-based predictions, the BFs depend on the Cabibbo-Kobayashi-Maskawa~(CKM) matrix elements $|V_{cs}|$ or $|V_{cd}|$ and on the nonperturbative transition form factors $F(q^2)$, where $q^2$ is the squared momentum transfer~\cite{4_theo_2_QCDSR,5_theo_3_CLFQM_2008,6_theo_4_CLFQM_2024,7_theo_6_CCQM_semionly,8_theo_5_BSW,9_theo_7_BSmethod,10_theo_8_Spin_symmetry}. Detailed theoretical predictions for $J/\psi$ semileptonic decays are summarized in Table~\ref{tab:theo_semi_leptonic}. The sum of the dominant $J/\psi$ semileptonic decay modes is predicted to be of order $10^{-9}$~\cite{4_theo_2_QCDSR, 5_theo_3_CLFQM_2008, 6_theo_4_CLFQM_2024}, which may be marginally observable at BESIII.
\begin{table*}[!htbp]
  \centering
  \caption{Theoretical predictions for BFs~(in units of $10^{-10}$) of $J/\psi$ semileptonic weak decays are presented. The transition form factors for $\psi \rightarrow D_{(s)}^{(*)-} l^{+} \nu_l$ are obtained using the ISGW model within the HQSS framework. For the other predictions, the transition form factors are computed with the corresponding models. Here, $\mathcal{B}(J / \psi \rightarrow D_{(s)}^{(*)-} l^{+} \nu_l)$ is summed over lepton flavors~($e$ and $\mu$) in the HQSS prediction. In BSW, the values based on the flavor-dependent average transverse quark momentum are quoted. For the theoretical predictions, only central values are reported.}
    \begin{tabular}{lcccccccc}
     \hline
    \hline
    Decay Channel & LQCD~\cite{3_theo_1_LQCD_semionly} & QCDSR~\cite{4_theo_2_QCDSR} & \multicolumn{2}{c}{CLFQM} & CCQM~\cite{7_theo_6_CCQM_semionly} & BSW~\cite{8_theo_5_BSW} & BS~\cite{9_theo_7_BSmethod}  & HQSS~\cite{10_theo_8_Spin_symmetry} \\
        &   &   &(2008)~\cite{5_theo_3_CLFQM_2008}   &(2024)~\cite{6_theo_4_CLFQM_2024}   &   &   &   &\\
    \hline
       $J / \psi \rightarrow D^{-} e^{+} \nu_e$   &   0.121(11)    &    0.073   &   0.51$\sim$0.57    &   0.610    &     0.171     &   0.60    & 0.203  & \multirow{2}[0]{*}{1.4} \\
        $J / \psi \rightarrow D^{-} \mu^{+} \nu_\mu$   &   0.118(11)    &   0.071   &   0.47$\sim$0.55    &   0.578    &     0.166     &   0.58    & 0.198  &  \\
        $J / \psi \rightarrow D_s^{-} e^{+} \nu_e$ &    1.90(8)   &   1.8   &   5.3$\sim$5.8    &   10.21    &   3.3    &   10.4    &   3.67    &  \multirow{2}[0]{*}{26.0}\\
        $J / \psi \rightarrow D_s^{-} \mu^{+} \nu_\mu$ &    1.84(8)   &   1.7    &   5.5$\sim$5.7    &   9.59    &   3.2    &   9.93    &   3.54    &  \\
        $J / \psi \rightarrow D^{*-} e^{+} \nu_e$  &    -   &     0.37  &   -    &   -    &    0.30   &    -   &    0.440    & \multirow{2}[0]{*}{2.3}\\
        $J / \psi \rightarrow D^{*-} \mu^{+} \nu_\mu$  &    -   &     0.36  &   -    &   -    &    0.29   &    -   &    0.424    & \\
         $J / \psi \rightarrow D_s^{*-} e^{+} \nu_e$ &   -    &   5.6    &    -   &    -   &  5.0     &   -    &    7.08    & \multirow{2}[0]{*}{42.0}\\
        $J / \psi \rightarrow D_s^{*-} \mu^{+} \nu_\mu$ &   -    &   5.4    &    -   &    -   &  4.8     &   -    &    6.75    & \\
        $\psi(2S) \rightarrow D^{-} e^{+} \nu_e$ &   -    &   -    &    -   &    0.345   &  -     &   -    &    -    & -\\
        $\psi(2S) \rightarrow D^{-} \mu^{+} \nu_\mu$ &   -    &   -    &    -   &    0.339   &  -     &   -    &    -    & -\\
        $\psi(2S) \rightarrow D_s^{-} e^{+} \nu_e$ &   -    &   -    &    -   &    7.20   &  -     &   -    &    -    & -\\
        $\psi(2S) \rightarrow D_s^{-} \mu^{+} \nu_\mu$ &   -    &   -    &    -   &    7.02   &  -     &   -    &    -    & -\\

    \hline
    \hline
    \end{tabular}
  \label{tab:theo_semi_leptonic}
\end{table*}

Notably, the ratio of Cabibbo-favored to Cabibbo-suppressed decays can be cleanly extracted from measurements of charmonium semileptonic decays, as many theoretical uncertainties cancel. This ratio therefore provides a clean probe of the effects of $\text{SU}(3)$ symmetry breaking. The ratios $R_{s/d}^l(\psi) \equiv \mathcal{B}\left( \psi \rightarrow D_s^{-} l^{+} \nu\right)/\mathcal{B}\left( \psi \rightarrow D^{-} l^{+} \nu\right)$ and $R_{s / d}^{*l}(\psi) \equiv \mathcal{B}\left( \psi \rightarrow D_s^{*-} l^{+} \nu\right)/\mathcal{B}\left(\psi \rightarrow D^{*-} l^{+} \nu\right)$ are expected to be $\left|V_{c s} / V_{c d}\right|^2\approx 19.46$ in the $\text{SU}(3)$ flavor-symmetry limit~\cite{6_theo_4_CLFQM_2024}. According to the PDG, the corresponding CKM matrix elements are $\left|V_{c s}\right|=0.975 \pm 0.006$ and $\left|V_{c d}\right|=0.221 \pm 0.004$~\cite{19_rev_5_pdg}. The ratios $R^{(*)l}_{s/d}$ predicted by various models are shown below. In CLFQM~\cite{6_theo_4_CLFQM_2024}, the predictions for electrons and muons are consistent within uncertainties:
\begin{equation}
    \begin{gathered}
        R^{e}_{s/d}(J/\psi) = 16.74 \pm 2.37,\\
        R^{\mu}_{s/d}(J/\psi) = 16.59 \pm 2.36,\\
        R^{e}_{s/d}(\psi(2S)) = 20.87 \pm 4.09,\\
        R^{\mu}_{s/d}(\psi(2S)) = 20.71 \pm 3.62,
    \end{gathered}
\end{equation}
and the prediction based on the CCQM~\cite{7_theo_6_CCQM_semionly} yields
\begin{equation}
    \begin{gathered}
        R^{l}_{s/d}(J/\psi) \approx 19.3,\\
        R^{*l}_{s/d}(J/\psi) \approx 16.6,
    \end{gathered}
\end{equation}
whereas the predictions from QCDSR~\cite{4_theo_2_QCDSR} are 
\begin{equation}
    \begin{gathered}
        R^{l}_{s/d}(J/\psi) \approx 24.7,\\
        R^{*l}_{s/d}(J/\psi) \approx 15.1.
    \end{gathered}
\end{equation}
These calculations suggest the presence of $\text{SU}(3)$ symmetry-breaking effects in charmonium semileptonic decays.

Another notable point is the ratios of BFs for decays involving $\mu$ and $e$, defined as $R_{J/\psi}(D/D_s) \equiv \frac{\mathcal{B}(J/\psi \to D/D_s\mu \nu_\mu) }{\mathcal{B}(J/\psi \to D/D_s e \nu_e)} $, since such ratios can serve as probes of lepton-flavor universality. According to lattice-QCD (LQCD) calculations, $R_{J/\psi}\left(D_s\right)=0.97002(8)$ and $R_{J / \psi}(D)=0.97423(15)$~\cite{3_theo_1_LQCD_semionly}, which await experimental verification with more precise measurements.

\subsection{Charmonium weak hadronic decays}

\begin{figure}[htbp]
    \centering
    \includegraphics[width=1.0\linewidth]{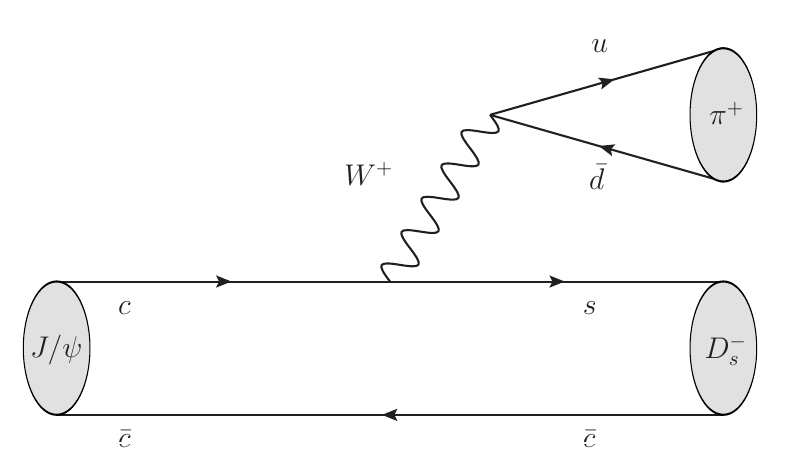}\label{fig:non-leptonic-sub1}
    \caption{The tree-level Feynman diagram for the charmonium nonleptonic weak decay $J/\psi \to D_{s}^{-} \pi^{+}$.
    }
    \label{fig:Feynman_diag_non-leptonic}
\end{figure}

Another category of charmonium weak decays comprises nonleptonic modes such as $\psi \to D_{(s)}^{(*)} + M$, where $M$ denotes a light meson~(e.g., $\pi$ or $\rho$), as shown in Fig.~\ref{fig:Feynman_diag_non-leptonic}. These decays are also $W$-mediated~($c \to s/d + u + \bar{d}/s$) but additionally involve a non-perturbative component: the hadronization into the light meson $M$. Analogous to the models introduced in the semileptonic decay section, theoretical calculations have provided predictions for hadronic weak decays using various frameworks, including the CLFQM~\cite{5_theo_3_CLFQM_2008,6_theo_4_CLFQM_2024}, QCDSR~\cite{13_theo_11_QCDSR_non-lep}, the BSW model~\cite{8_theo_5_BSW}, the BS method~\cite{9_theo_7_BSmethod}, HQSS~\cite{10_theo_8_Spin_symmetry}, and the factorization approximation~\cite{11_theo_9_Spin_symmetry, 14_theo_12_factorization_approximation_non-lep}.

Based on their final states, charmonium nonleptonic two-body weak decays can be classified as $\psi\rightarrow PP/PV/VV$, where $P$ and $V$ denote pseudoscalar and vector mesons, respectively. For the mode $\psi\rightarrow PP$, the Cabibbo-favored, color-allowed process $J/\psi \rightarrow D_s^{-} \pi^{+}$ dominates. Table~\ref{tab:theo_non_leptonic_PP} lists the predicted BFs for the $\psi\rightarrow PP$ mode. 

For the mode $\psi\rightarrow PV$, Table~\ref{tab:theo_non_leptonic_PV} lists the predicted BFs. Among all channels, the Cabibbo-favored, color-allowed process $\psi \rightarrow D_s^{-} \rho^{+}$ is dominant.

For the mode $\psi\rightarrow VV$ listed in Table~\ref{tab:theo_non_leptonic_VV}, the most accessible decay is $J/\psi \rightarrow D_s^{*+} \rho^{-}$, which has a predicted branching fraction of $5.26\times10^{-9}$ from QCDSR~\cite{13_theo_11_QCDSR_non-lep} and $5.86\times10^{-9}$ from the QCD factorization approach~\cite{14_theo_12_factorization_approximation_non-lep}. This decay mode has the highest probability of being observed in the future.

Several BSM scenarios can enhance these branching fractions. For example, in the decay $J/\psi \rightarrow \bar{D}^0 X_u$ (where $X_u$ denotes a meson containing a $u$ quark, such as $\pi^\pm$, $\pi^0$, or $\rho$), an estimate by A. Datta et al. suggests that flavor-changing processes arising from the TopColor model or two-Higgs-doublet models (2HDM) could yield branching fractions in the range $(0.1\text{--}2.0) \times 10^{-5}$ in the TopColor scenario (for a top-pion mass $m_{\tilde{\pi}}~\text{between}~100\text{ and }200~\text{GeV}/c^2$ with mixing angles $\sim 1$) and $(0.1\text{--}1.4) \times 10^{-7}$ in the 2HDM scenario (for a 2HDM Higgs mass $m_H~\text{between}~100\text{ and }200~\text{GeV}/c^2$ with all couplings $\sim 1$)~\cite{17_BSM_1}.

\begin{table*}[!htbp]
  \centering
  \caption{Predictions for the branching fractions~(in units of $10^{-10}$) of $J/\psi\rightarrow PP$ decays. In the BSW model, the values based on the flavor-dependent average transverse quark momentum $\omega$ are quoted. For the theoretical predictions, only the central values are reported.}
    \begin{tabular}{llcccccc}
     \hline
    \hline
     Transition Mode & Decay Channel  & QCDSR~\cite{13_theo_11_QCDSR_non-lep} & CLFQM  & BSW~\cite{8_theo_5_BSW} & BS~\cite{9_theo_7_BSmethod}  & HQSS~\cite{11_theo_9_Spin_symmetry} & Factorization~\cite{14_theo_12_factorization_approximation_non-lep} \\
        &   &   &(2024)~\cite{6_theo_4_CLFQM_2024}   &   &   &   &\\
    \hline
       \multirow{2}[0]{*}{$\Delta C=\Delta S=+1$}   &   $J/\psi \rightarrow D_s^{-} \pi^{+}$ &  2.0    &   3.64    &     7.41     &   4.75    & 8.74  & 10.9\\
       &  $J/\psi \rightarrow \bar{D}^0 \bar{K}^0$& 0.36 & -  & 1.39  & 0.803  & 2.80  &  1.44 \\
       \hline
       \multirow{5}[0]{*}{$\Delta C=+1,\Delta S=0$}&  $J/\psi \rightarrow D_s^{-} K^{+}$& 0.16  & 0.202  &  0.53  & 0.312  &  0.55 &    0.618  \\
       &  $J/\psi \rightarrow D^{-} \pi^{+}$& 0.080 & 0.190  & 0.29  &  0.183 & 0.55  & 0.637\\
       &  $J/\psi \rightarrow \bar{D}^0 \pi^0$  & -  & - & 0.024  & 0.0156  & 0.055  &  0.0350 \\
       &  $J/\psi \rightarrow \bar{D}^0 \eta$ &  -  & - & 0.070  &  0.00263 & 0.016  &   0.0103\\
       &  $J/\psi \rightarrow \bar{D}^0 \eta^{\prime}$  &  - & -  &  0.004 & 0.0371  & 0.003  &   0.00583\\
       \hline
       \multirow{2}[0]{*}{$\Delta C=+1,\Delta S=-1$}&  $J / \psi \rightarrow D^{-} K^{+}$ &  -  & 0.0116  &  0.023  & 0.0131  & -  &  0.0379 \\
       &  $J / \psi \rightarrow \bar{D}^0 K^0$ & -  & -  &  0.004  & 0.00224  & -  &  0.00416 \\

    \hline
    \hline
    \end{tabular}
  \label{tab:theo_non_leptonic_PP}
\end{table*}

\begin{table*}[!htbp]
  \centering
  \caption{Predictions for BFs (in units of $10^{-10}$) of $J/\psi\rightarrow PV$ decays. In BSW, values based on the flavor-dependent average transverse quark momentum, $\omega$, are quoted. For the theoretical predictions, only the central values are quoted.}
    \begin{tabular}{llcccccc}
     \hline
    \hline
     Transition Mode & Decay Channel  & QCDSR~\cite{13_theo_11_QCDSR_non-lep} & CLFQM  & BSW~\cite{8_theo_5_BSW} & BS~\cite{9_theo_7_BSmethod}  & HQSS~\cite{11_theo_9_Spin_symmetry} & Factorization~\cite{14_theo_12_factorization_approximation_non-lep} \\
        &   &   &(2024)~\cite{6_theo_4_CLFQM_2024}   &   &   &   &\\
    \hline
       \multirow{2}[0]{*}{$\Delta C=\Delta S=+1$}   &   $J/\psi \rightarrow D_s^{-} \rho^{+}$ &  12.6    &   29.5    &     51.1     &   26.2    & 36.30  & 38.2\\
       &  $J/\psi \rightarrow \bar{D}^0 \bar{K}^{*0}$& 1.54 & -  & 7.61  & 4.75  & 10.27  &  4.09 \\
       \hline
       \multirow{5}[0]{*}{$\Delta C=+1,\Delta S=0$}&  $J/\psi \rightarrow D_s^{-} K^{*+}$& 0.82  & 1.42  &  2.82  & 1.67  &  2.12 &    2.00  \\
       &  $J/\psi \rightarrow D^{-} \rho^{+}$& 0.42 & 1.70  & 2.16  &  1.13 & 2.20  & 2.12\\
       &  $J/\psi \rightarrow \bar{D}^0 \rho^0$  & -  & - & 0.18  & 0.0960  & 0.22  &  0.108 \\
       &  $J/\psi \rightarrow \bar{D}^0 \omega$ &  -  & - & 0.16  &  0.0880 & 0.18  &   0.0810\\
       &  $J/\psi \rightarrow \bar{D}^0 \phi$  &  - & -  &  0.42 & 0.307  & 0.65  &   0.192\\
       \hline
       \multirow{2}[0]{*}{$\Delta C=+1,\Delta S=-1$}&  $J / \psi \rightarrow D^{-} K^{*+}$ &  -  & 0.0859  &  0.13  & 0.0770  & -  &  0.114 \\
       &  $J / \psi \rightarrow \bar{D}^0 K^{*0}$ & -  & -  &  0.021  & 0.0132  & -  &  0.0119 \\

    \hline
    \hline
    \end{tabular}
  \label{tab:theo_non_leptonic_PV}
\end{table*}

\begin{table}[!htbp]
  \centering
  \caption{Predictions for the BFs~(in units of $10^{-10}$) of $J/\psi\rightarrow VV$ decays. For the theoretical predictions, only the central values are quoted.}
    \begin{tabular}{lccc}
     \hline
    \hline
    Decay Channel  & QCDSR~\cite{13_theo_11_QCDSR_non-lep} & BS~\cite{9_theo_7_BSmethod} \\
    \hline
        $J / \psi \rightarrow D_s^{*-} \rho^{+}$ &   52.6   &   58.6 \\
        $J / \psi \rightarrow D_s^{*-} K^{*+}$& 2.6  & 2.62  \\
        $J / \psi \rightarrow D^{*-} \rho^{+}$ & 2.8  & 3.30  \\
       $J / \psi \rightarrow \bar{D}^{*0} \bar{K}^{*0}$ & 9.6   &  11.1 \\
    \hline
    \hline
    \end{tabular}
  \label{tab:theo_non_leptonic_VV}
\end{table}

\subsection{Charmonium FCNC decays}
\begin{figure}[htbp]
    \centering
    \includegraphics[width=1.0\linewidth]{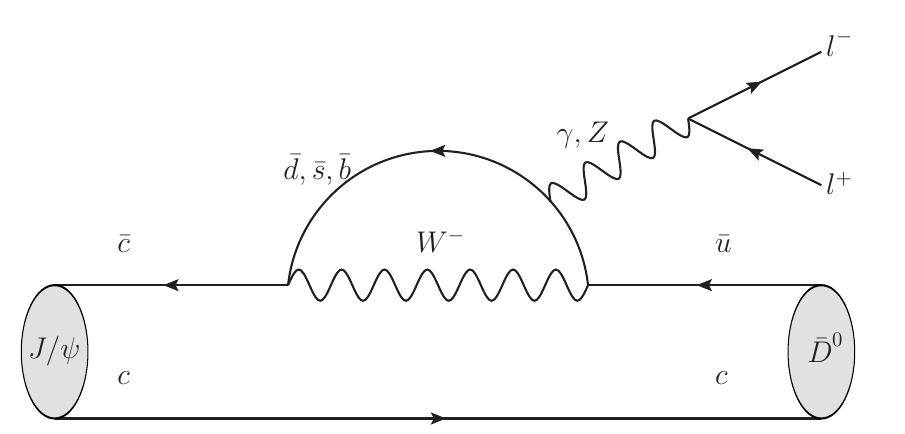}
    \caption{Feynman diagram for the charmonium FCNC decay $J/\psi \to \bar{D}^0 l^+ l^-$. 
    }
    \label{fig:Feynman_Diag_FCNC}
\end{figure}

Beyond the aforementioned tree-level SM processes sensitive to NP, the FCNC decays $\psi \to \bar{D}^0 l^+ l^-$, shown in Fig.~\ref{fig:Feynman_Diag_FCNC}, are forbidden at tree level by the Glashow-Iliopoulos-Maiani~(GIM) mechanism and are highly suppressed at loop level~\cite{1_rev_1}. Theoretical predictions based on QCDSR~\cite{15_theo_13_FCNC_QCDSR} are summarized in Table~\ref{tab:theo_FCNC}. Given the extremely small BFs of these FCNC decays, any observation of such processes at BESIII would provide unambiguous evidence of NP, potentially arising from TopColor models~\cite{20_BSM_2_TopColor}, the minimal supersymmetric standard model~\cite{21_BSM_3_MSSM}, and the 2HDM~\cite{22_BSM_4_2HDM}.

\begin{table}[!htbp]
  \centering
  \caption{Predictions for the BFs~(in units of $10^{-13}$) of the FCNC decay $J/\psi\rightarrow \bar{D}^0l^+l^-$. For the theoretical predictions, only the central values are quoted.}
    \begin{tabular}{lccc}
     \hline
    \hline
    Decay Channel  & QCDSR~\cite{15_theo_13_FCNC_QCDSR} \\
    \hline
        $J / \psi \rightarrow \bar{D}^0 e^{+} e^{-}$ &   1.14    \\
        $J / \psi \rightarrow \bar{D}^{* 0} e^{+} e^{-}$ & 6.30   \\
        $J / \psi \rightarrow \bar{D}^0 \mu^{+} \mu^{-}$ & 1.08   \\
       $J / \psi \rightarrow \bar{D}^{* 0} \mu^{+} \mu^{-}$ & 5.94 \\
    \hline
    \hline
    \end{tabular}
  \label{tab:theo_FCNC}
\end{table}

\section{Experimental Searches at BESIII}

\begin{table*}[!htbp]
    \centering
    \caption{Summary of upper limits from charmonium weak-decay searches at BESIII.}
    \label{tab:summary}
    \begin{tabular}{l c c c}
        \hline
        \hline
        Decay Channel & $J/\psi,\psi(2S)$ events~($\times10^6$) &  Measured Upper Limit  & SM Prediction\\
        \hline
        $J/\psi \to D^{-}e^{+}\nu_{e} $ & $10087$ & $< 7.1 \times 10^{-8}$~\cite{18_rev_4}  & $\sim 10^{-11}$\\
        $J/\psi \to D^{-} \mu^{+}\nu_{\mu} $ & $10087$ & $< 5.6 \times 10^{-7}$~\cite{23_exp_1} & $\sim 10^{-11}$\\ 
        $J/\psi \to D_{s}^{-}e^{+}\nu_{e} $ & $10087$ & $< 9.9 \times 10^{-8}$~\cite{24_exp_2} & $\sim 10^{-10}$ \\
        $J/\psi \to D_{s}^{*-}e^{+}\nu_{e} $ & $225$ & $< 1.8 \times 10^{-6}$~\cite{24_exp_2_1} & $\sim 10^{-10}$\\       
        $J/\psi \to D_{s}^{-}\rho^{+} $ & $10087$ & $< 8.0 \times 10^{-7}$~\cite{25_exp_3} & $\sim 10^{-9}$\\        
        $J/\psi \to D_{s}^{-}\pi^{+} $ & $10087$ & $< 4.1 \times 10^{-7}$~\cite{25_exp_3} & $\sim 10^{-10}$\\        
        $J/\psi \to D^{-} \pi^{+} $ & $10087$ & $< 7.0 \times 10^{-8}$~\cite{26_rev_6} &$\sim 10^{-11}$\\        
        $J/\psi \to D^{-} \rho^{+} $ & $10087$ & $< 6.0 \times 10^{-7}$~\cite{26_rev_6} & $\sim 10^{-10}$\\        
        $J/\psi \to \bar{D}^{0} \pi^{0} $ & $10087$ & $< 4.7 \times 10^{-7}$~\cite{26_rev_6} &$\sim 10^{-12}$\\       
        $J/\psi \to \bar{D}^{0} \eta $ & $10087$ & $< 6.8 \times 10^{-7}$~\cite{26_rev_6} & $\sim 10^{-12}$ \\      
        $J/\psi \to \bar{D}^{0} \rho^{0} $ & $10087$ & $< 5.2 \times 10^{-7}$~\cite{26_rev_6} &$\sim 10^{-11}$\\        
        $J/\psi \to \bar{D}^{0}\bar{K}^{*0} $ & $10087$ & $< 1.9 \times 10^{-7}$~\cite{27_exp_7} &$\sim 10^{-10}$\\           
        $J/\psi \to D^{0}\mu^{+}\mu^{-} $ & $10087$ & $< 1.1 \times 10^{-7}$~\cite{28_exp_8} & $\sim 10^{-13}$\\      
        $J/\psi \to D^{0}e^{+}e^{-} $ & $1311$ & $< 8.5 \times 10^{-8}$~\cite{29_exp_9} & $\sim 10^{-13}$\\        
        $J/\psi \to \gamma D^{0} $ & $10087$ & $< 9.1 \times 10^{-8}$~\cite{30_exp_10} & $\sim 10^{-13}$\\
        $\psi(2S) \to D^{0}e^{+}e^{-} $ & $448$ & $< 1.4 \times 10^{-7}$~\cite{29_exp_9} & $\sim 10^{-13}$\\
        $\psi(2S) \to \Lambda_{c}^{+}\bar{p}e^+e^-$ & $448$ & $< 1.7 \times 10^{-6}$~\cite{33_exp_12} & $\sim 10^{-10}$\\
        $\psi(2S) \to \Lambda_{c}^{+}\bar{\Sigma}^{-} $ & $448$ & $< 1.4 \times 10^{-5}$~\cite{32_exp_11} & $\sim 10^{-10}$\\
        \hline
        \hline
    \end{tabular}
      \label{tab:exp}
\end{table*}

\subsection{Charmonium semileptonic decays}

Semileptonic decays are crucial because they involve both CKM matrix elements and nonperturbative QCD form factors, providing a unique platform for testing fundamental interactions and bound-state dynamics~\cite{6_theo_4_CLFQM_2024}. The upper limits presented later are all at the 90\% confidence level~(\text{C.L.}).

$\boldsymbol{J/\psi \rightarrow D \ell \nu_{\ell}}$ \textbf{channels}: The BESIII Collaboration has performed dedicated searches for the semileptonic decays of the $J/\psi$ to light charmed mesons~($D$) and strange charmed mesons~($D_s$). The search for $J/\psi \to D^{-}e^{+}\nu_{e}$~(Cabibbo-suppressed, with $|V_{cd}|$ dependence) used a sample of $10.1 \times 10^{9}$ $J/\psi$ events, with $D^{-}$ reconstructed via $D^{-} \to K^+\pi^-\pi^-$. No significant signal was observed, allowing the collaboration to set the most stringent upper limit on the branching fraction to date.
\begin{equation}
\mathcal{B}(J/\psi \to D^{-}e^{+}\nu_{e} ) < 7.1 \times 10^{-8}~\text{\cite{18_rev_4}}.  
\end{equation}

A search for $J/\psi \to D^{-}\mu^{+}\nu_{\mu} $~(Cabibbo-suppressed) was also conducted, marking the first investigation of a weak charmonium decay involving a muon in the final state. This search used the same sample of $10.1 \times 10^{9}$ $J/\psi$ events. As in the electron channel, no significant signal was observed, and an upper limit was set.
\begin{equation}
\mathcal{B}(J/\psi \to D^{-}\mu^{+}\nu_{\mu} ) < 5.6 \times 10^{-7}~\text{\cite{23_exp_1}}.    
\end{equation}

$\boldsymbol{J/\psi \to D_s \ell \nu_{\ell}}$ \textbf{channels}: Decays involving the $D_s$ meson are governed by the Cabibbo-favored CKM element $|V_{cs}|$. BESIII searched for $J/\psi~\to~D_s^- e^+ \nu_e$ and the corresponding vector-meson channel $J/\psi~\to~D_s^{*-} e^+ \nu_e$, collectively denoted as $J/\psi \to D_{s}^{(*)-}e^{+}\nu_{e}$. Using the full data set of $10.1 \times 10^{9}$ $J/\psi$ events, BESIII set upper limits on these decays. 
\begin{equation}
\mathcal{B}(J/\psi \to D_{s}^{-}e^{+}\nu_{e} ) < 9.9 \times 10^{-8}~\text{\cite{24_exp_2}}.  
\end{equation}
By contrast, for the $J/\psi \to D_{s}^{*-}e^{+}\nu_{e}$ decay, the published upper limit was set using only $2.25 \times 10^8$ $J/\psi$ events.  
\begin{equation}
\mathcal{B}(J/\psi \to D_{s}^{*-}e^{+}\nu_{e} ) < 1.8 \times 10^{-6}~\text{\cite{24_exp_2_1}}.    
\end{equation}

In these analyses, the $D_{s}^{-}$ mesons are reconstructed in four decay modes: $D_{s}^{-} \to K_s^0 K^-$, $D_{s}^{-} \to K^+K^-\pi^-$, $D_{s}^{-} \to K^+K^-\pi^-\pi^0$, and $D_{s}^{-} \to K_s^0 K^- \pi^+ \pi^-$. The $D_{s}^{*-}$ mesons are reconstructed via $D_{s}^{*-} \to D_{s}^{-} \gamma$. The ongoing analysis with the full $J/\psi$ dataset will push the sensitivity of these semileptonic channels closer to the precise SM predictions, enabling stringent tests of theoretical models for transition form factors.

\subsection{Charmonium weak hadronic decays}

Nonleptonic weak decays, $J/\psi \to D M$, are particularly challenging due to the entirely hadronic nature of the final state, which introduces large theoretical uncertainties related to the hadronic matrix elements~\cite{13_theo_11_QCDSR_non-lep}. Beyond these uncertainties, the nonleptonic final-state mesons $D$ and $M$ both predominantly decay into light hadrons. Since these light-hadron final states are usually identical to those from the dominant strong decay modes of $J/\psi$, a full reconstruction of the nonleptonic decay faces overwhelming backgrounds. Therefore, such measurements are typically performed by tagging the signal via a semileptonic decay of the $D$ meson.

$\boldsymbol{J/\psi \to D_s M}$ \textbf{channels}: Decays into a charmed-strange meson $D_s$ and a light meson $M$ are Cabibbo-favored. BESIII recently conducted searches for $J/\psi \to D_{s}^{-}\rho^{+}$ and $J/\psi \to D_{s}^{-}\pi^{+}$ using the full $J/\psi$ data set, in which the semileptonic decay mode $D_{s}^{-}~\to~\phi(\to~K^+K^-) e^{-} \bar{\nu}_e~$ was used to tag the signal. No significant signal was observed in either channel. The resulting upper limits are the most stringent constraints to date: 
\begin{equation}
    \mathcal{B}(J/\psi \to D_{s}^{-}\rho^{+} ) < 8.0 \times 10^{-7}~\text{\cite{25_exp_3}},
\end{equation}
\begin{equation}
    \mathcal{B}(J/\psi \to D_{s}^{-}\pi^{+} ) < 4.1 \times 10^{-7}~\text{\cite{25_exp_3}}.
\end{equation}

$\boldsymbol{J/\psi \to D M}$ \textbf{channels}: BESIII has also performed comprehensive searches for Cabibbo-suppressed two-body nonleptonic decays that depend on the CKM element $|V_{cd}|$. Using the full $J/\psi$ data set and tagging $D^{-}$ with $D^{-} \to K_s^0 e^{-} \bar{\nu}_e$ and $\bar{D}^{0}$ with $\bar{D}^{0} \to K^+ e^- \bar{\nu}_e$, BESIII has established the following upper limits for final states containing a nonstrange $D$ meson:
\begin{equation}
    \mathcal{B}(J/\psi \to D^{-} \pi^{+} ) < 7.0 \times 10^{-8}~\text{\cite{26_rev_6}},
\end{equation}
\begin{equation}
    \mathcal{B}(J/\psi \to D^{-} \rho^{+} ) < 6.0 \times 10^{-7}~\text{\cite{26_rev_6}},
\end{equation}
\begin{equation}
    \mathcal{B}(J/\psi \to \bar{D}^{0} \pi^{0} ) < 4.7 \times 10^{-7}~\text{\cite{26_rev_6}},
\end{equation}
\begin{equation}
\mathcal{B}(J/\psi \to \bar{D}^{0} \eta ) < 6.8 \times 10^{-7}~\text{\cite{26_rev_6}},
\end{equation}
\begin{equation}
\mathcal{B}(J/\psi \to \bar{D}^{0} \rho^{0} ) < 5.2 \times 10^{-7}~\text{\cite{26_rev_6}},
\end{equation}
\begin{equation}
\mathcal{B}(J/\psi \to \bar{D}^{0}\bar{K}^{*0} ) < 1.9 \times 10^{-7}~\text{\cite{27_exp_7}}.
\end{equation}

\subsection{Charmonium FCNC decays}

Flavor-changing neutral current decays in the charm sector are highly suppressed by the GIM mechanism. BESIII has searched for FCNC decays with both muon- and electron-pair final states. Using a sample of $(10087\pm44)\times10^{6}$ $J/\psi$ events, BESIII has set upper limits of 
\begin{equation}
\mathcal{B}(J/\psi \to D^{0}\mu^{+}\mu^{-} ) < 1.1 \times 10^{-7}~\text{\cite{28_exp_8}},
\end{equation}
\begin{equation}
    \mathcal{B}\left(J / \psi \rightarrow \gamma D^0\right)<9.1 \times 10^{-8}~\text{\cite{30_exp_10}}.
\end{equation}
Using $1.3\times10^{9}$ $J/\psi$ events, BESIII set an upper limit on $J/\psi\rightarrow D^0 e^+e^-$. 
\begin{equation}
    \mathcal{B}(J/\psi \to D^{0}e^{+}e^{-} ) < 8.5 \times 10^{-8}~\text{\cite{29_exp_9}}.
\end{equation}

In these analyses, the $D^{0}$ candidates are reconstructed in three decay modes: $D^0 \to K^-\pi^+$, $D^0 \to K^-\pi^+\pi^0$, and $D^0 \to K^-\pi^+\pi^+\pi^-$.
An update of the result based on $10.1 \times 10^{9}$ $J/\psi$ events is in progress.
These world-leading limits constrain NP models that allow FCNC transitions in the heavy quarkonium system.

\subsection{Weak Decays of $\psi(2S)$}

Searches for weak decays have been extended to the $\psi(2S)$ state, which features a similar hadronic structure but a larger mass and different decay channels, thereby offering access to unique final states and sensitivity to distinct new-physics scenarios.

A search for the FCNC decay $\psi(2S) \to D^{0}e^{+}e^{-} $ yielded an upper limit on the branching fraction. 
\begin{equation}
\mathcal{B}(\psi(2S) \to D^{0}e^{+}e^{-} ) < 1.4 \times 10^{-7}~\text{\cite{29_exp_9}}.
\end{equation}
In addition, a search for the decay $\psi(2 S) \rightarrow \Lambda_c^{+} \bar{p} e^{+} e^{-}$ was performed, yielding an upper limit of
\begin{equation}
\mathcal{B}(\psi(2S) \to \Lambda_{c}^{+}\bar{p}e^+e^-) < 1.7 \times 10^{-6}~\text{\cite{33_exp_12}}.
\end{equation}

Furthermore, a search has been performed for the weak, baryonic decay $\psi(2S) \to \Lambda_{c}^{+}\overline{\Sigma}^{-}$, which is theoretically predicted to have a branching fraction of approximately $10^{-10}$~\cite{31_theo_14}. This is the first search for a purely baryonic weak decay of the $\psi(2S)$ and yields an upper limit. 
\begin{equation}
\mathcal{B}(\psi(2S) \to \Lambda_{c}^{+}\bar{\Sigma}^{-} ) < 1.4 \times 10^{-5}~\text{\cite{32_exp_11}}.
\end{equation}

At present, all upper limits on $\psi(2S)$ weak decays have been set using a sample of $448\times10^6$ $\psi(2S)$ events, and new searches using the full data set of $2.7\times10^9$ $\psi(2S)$ events are in progress.

\section{Summary and Outlook}

Studies of the weak decays of heavy quarkonia such as $J/\psi$ and $\psi(2S)$ serve as ideal probes of non-perturbative QCD effects and $\text{SU}(3)$ symmetry breaking effects. The BESIII experiment has been extremely successful in utilizing its large $J/\psi$ and $\psi(2S)$ datasets to probe the weak decay sector of charmonium. The experimental results, as summarized in Table~\ref{tab:exp}, have provided the most stringent upper limits on these rare processes to date.

The current upper limits for semileptonic and nonleptonic decays remain above the highest SM predictions. For limits derived using only a fraction of the total $\psi(2S)$ or $J/\psi$ datasets, updates using the full data samples are underway. Future data collected at BESIII and at prospective higher-luminosity facilities~(e.g., the Super Tau-Charm Facility~(STCF)\cite{STCF_ref}) offer the potential to push the upper limits down to $10^{-9}$ or even lower and eventually observe the SM weak decays of charmonium. The theoretical community has provided the necessary form factor calculations using LQCD, QCDSR, and other non-perturbative methods, setting the stage for direct comparison with future experimental observations.

Crucially, searches for FCNC processes, such as $J/\psi \to D^0 \mu^+ \mu^-$ and $J/\psi \to D^0 e^+ e^-$, have placed strong constraints on BSM theories, limiting the parameter space for new particles and interactions that couple to the charm sector. The pursuit of charmonium weak decays remains a key component of the BESIII physics program and a dynamic field full of challenges and opportunities for both theory and experiment.

\acknowledgments
\hspace{1.5em}
This work was supported in part by the National Key R\&D Program of China under Grant No. 2023YFA1606000; the National Natural Science Foundation of China (NSFC) under Grant Nos. 12035009, 12175321, and U1932101; and the National College Students' Science and Technology Innovation Project of Sun Yat-sen University.

\section*{References}

\bibliography{refs}
\end{document}